\documentclass[12pt]{iopart}
\usepackage{iopams}  
\usepackage{graphicx}% Include figure files
\usepackage{dcolumn}% Align table columns on decimal point
\usepackage{bm}% bold math
\usepackage{epsfig}

\begin{document}
% Journal identifier can be put here if required, e.g.
%\jl{14}

%=============================================================== 
\title{Ripples and Ripples: from Sandy Deserts to Ion-Sputtered Surfaces}
%\subtitle{}
%=============================================================== 

\author{T. Aste$^1$ and U. Valbusa$^2$}
%\author{  }
%\email{tomaso.aste@anu.edu.au }
\address{$^1$ Applied Mathematics, RSPhysSE, ANU, 0200 Canberra ACT, Australia }

\address{$^1$ INFM-Dipartimento di Fisica, Universit\`a di Genova, via Dodecaneso 33, 16146 Genova, Italy}

\date{12/04/04}                          % change for final

%\date{Received: date / Revised version: date}
%
\begin{abstract}
We study the morphological evolution of surfaces during ion sputtering and we compare their dynamical roughening with aeolian ripple formation in sandy deserts. 
We show that, although the two phenomena are physically different, they must obey to similar geometrical constraints and therefore can be described within the same theoretical framework. 
The present theory distinguish between atoms that stay bounded in the bulk and others that are mobile on the surface.
We describe the excavation mechanisms, the adsorption and the surface mobility by means of a continuous equation derived from the study of dune formation on sand.
This approach can explain the different dynamical behaviors experimentally observed in metals or in semiconductors and amorphous systems. 
We also show that this novel approach can describe the occurrence of ripple rotation in the $(x,y)$ plane induced by changes in the sputtering incidence angle. 
\end{abstract}
\pacs{
	%{PACS-key} {describing text of that key} \and
	 {45.70.-n} %{Granular systems} \and
	 {61.43.-j} %{Disordered Solids} \and 
	 {68.35.Ct} %{Interface structure and roughness} \and
	 {68.55.Jk} %{Structure and morphology}. 
} % end of PACS codes

\maketitle

\section{Introduction}

When an ion hits a surface liberates locally a large amount of energy that melts a region of the solid immediately below.
For geometrical reasons, the sputtering effect depends on the surface-curvature: the energy concentrates on regions of positive curvature and this favorites the excavation of valleys and the growth of hills.
On the other hand, thermal diffusion and surface tension tend to smoothen the irregularities by flattening the surface.
It has been observed that under the combined action of these mechanisms the surface tends to create spontaneously ripples \cite{Ekl91,Krim93,Chra94,Valbusa02,Buat00,Rusp98,Rusp98a,Rusp97}. 
In nature, ripples are commonly observed in sandy-deserts as the result of dynamical instability of the sand surface under the action of a sufficiently strong wind \cite{Bagnold}.
In this case, the formation of ripples is commonly associated with the effect produced by some grains that are lifted from the sand-bed and accelerated by the wind.
These grains, when re-impact with the bed, splash up a number of other grains. Most of these grains return to the bed leading to a local rearrangement, whereas some other are accelerated by the wind and impact again on the bed after a certain `saltation' length.
In the literature, many studies have been devoted to understanding the mechanism of ripple formation \cite{Andersen02,Andersen01,Kurtze00,Hoyle99,Prigozhin99,Werner93}.
In particular, an hydrodynamical model for aeolian ripple formation, based on a continuum dynamical description with two species of grains (immobile and rolling grains), was proposed with success by Bouchaud et al. \cite{Bouch95,Terz98,Val99,Csa00}. 
The main ingredient of such a model is a bilinear differential equation, for the population of the two species of grains, which shows the instability of a flat bed against ripple formation.

%%%%%%%%%%%%%%%%%%%%%%%%%%%%%%%%%%%%%%%%%%%%%%%%%%%%%%%%%%%%%%%%%%%
%%%%%%%%%%%%%%%%%%%%%%%%%%%%%%%%%%%%%%%%%%%%%%%%%%%%%%%%%%%%%%%%%%%
\begin{figure}
%\hspace{+1cm}
\begin{center}
\includegraphics[width=0.75\textwidth]{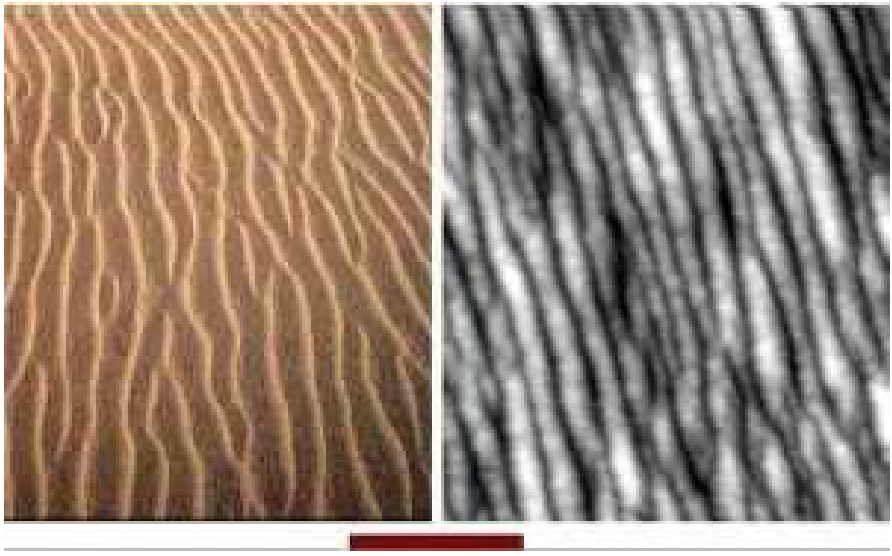}
\end{center}
\caption{\label{f.0} %\footnotesize
Ripples on sand (Gobi desert) and Ripples on surfaces (Ag under ion sputtering).
The line is indicative of the different length scales involved: it represents 1 $m$ in the left picture and 50 $nm$ in the right one. }
\end{figure}
%%%%%%%%%%%%%%%%%%%%%%%%%%%%%%%%%%%%%%%%%%%%%%%%%%%%%%%%%%%%%%%%%%%
%%%%%%%%%%%%%%%%%%%%%%%%%%%%%%%%%%%%%%%%%%%%%%%%%%%%%%%%%%%%%%%%%%%

In this paper we show that the same reasoning which has been used to describe the sand ripples formation in deserts, applied to the studies of dynamical surface roughening, leads to an accurate description of the morphogenesis and evolution of ripples on crystal and amorphous surfaces during ion sputtering.
The present approach contains the Bradley-Harper approach \cite{Bra88} and its nonlinear extension by Rost et al. \cite{Rost95} and by Park et al. \cite{Park99} (based on a Kuramoto-Sivashisky and Kardar-Parisi-Zhang type equations \cite{Kur84,KPZ}) and it is also able to describe some of the crucial experimental features observed in these systems \cite{Rusp98}.
In particular, by means of this approach we can describe the two distinct dynamical behaviors experimentally observed in amorphous/semiconductors systems and in metals \cite{Valbusa02}. 
In the first case (amorphous/semiconductors) we find that the ripples growth exponentially fast at constant wavelength $\hat \lambda$ up to a critical roughening $W_c$ at which the growing process interrupts.
On the other hand, in metals (when the Erlich-Schwoebel barrier is active), we can observe a transition between an initial exponential to a slower power-law growth of the roughness (the root mean square of the height profile).
In this regime the ripple-wavelength tends also to growth with time and re-nucleation effects are observed.
Finally, we show that the intriguing phenomena of ``ripple rotation'' \cite{Rusp98}, associated with a change in the sputtering incidence angle, can be conveniently described by means of the present approach.

The main novelty of this work is the use of a new kind of equation to describe the surface instabilities under ion sputtering.
In the present approach we clearly distinguish between three mechanisms: 1) Erosion; 2) Adsorption-Condensation; 3) Mobility. 
Physically, all these three mechanisms take place simultaneously at a surface under ion-bombardment. 
The possibility to distinguishing between them clarifies the approach, simplifies the choice of the parameters and enlarges the descriptive power of the theory. 
Moreover, we distinguish between atoms that stay in the bulk in bounded position and others that stay on the surface and are mobile. 
This (simplified) view is more realistic than considering the whole system as a fluid. 
We note that these novel elements of the present theory do not introduce -per se- extra parameters in the analogy with the well established Bradley-Harper theory \cite{Bra88}.
However, our choice in this paper has been to study the equations in their most complete form introducing therefore extra parameters.
We show that, from the linear analysis, the Bradley-Harper solution for the ripple wavelength as well as other solutions for typical ripple periodicity in sandy deserts can be all retrieved and extended.

\section{Particle mobility and Ripple dynamics}

When the surface of a solid is taken under ion sputtering some atoms in the proximity of the surface receive energy from the sputtered ions and pass from a bounded - \emph{`immobile'} - solid state to a \emph{`mobile'} melted state.
The opposite mechanism is also allowed: some \emph{mobile} atoms can gain in energy by becoming \emph{immobile} and bounding in a given position in the solid.
A certain fraction of atoms might also be dispersed into the atmosphere. 
Let us call $h({\mathbf r },t)$ the height of surface profile made of immobile -bounded- atoms and call $R({\mathbf r },t)$ the height of mobile -melted- atoms.
In analogy with the theory developed to explain the dynamical evolutions of dunes in deserts \cite{Bouch95,Terz98,Val99,Csa00}, we describe the mechanisms of excavation, exchange between mobile and immobile atoms and surface displacement of mobile atoms in term of the following differential equation:
\begin{eqnarray} \label{E1}
\frac{\partial h}{\partial t} 
&=&  - \Gamma(R,h)_{ex} + \Gamma(R,h)_{ad} 
\nonumber \\
\frac{\partial R}{\partial t} 
&=& - {\mathbf \nabla} {\mathbf J}(R,h) + (1-\phi) \Gamma(R,h)_{ex} -\Gamma(R,h)_{ad} \;\;\;.
\nonumber \\
\end{eqnarray}
Where $\Gamma(R,h)_{ex}$ is the rate of atoms that are excavated under the action of the sputtering, and $(1-\phi)$ is the part of them that pass from immobile to 
mobile, whereas $\phi$ is the fraction that is dispersed into the atmosphere. 

Let us now write in details the various terms contained in Eq.\ref{E1}.

\subsection{Excavation}
The excavation effect must clearly depend on the number and velocity of the sputtered ions (i.e. its flux), but also the local shape and orientation of the surface might play an important role.
Indeed, the energy transmitted by the impacting ions concentrate more in regions of the surface with positive curvature. 
Moreover, part of the surface facing the flux are likely to experience a different erosion respect to others which are less exposed to the flux.
Such dependence of the sputtering yields from the combination of the surface curvature and the incoming ion flux orientation has been described in great details in the literature (see for instance \cite{Valbusa02} and references therein). 
Here we adopt the same reasoning that lead to the Bradley and Harper equation \cite{Bra88}.
Crystalline orientation and anisotropies might be also taken into account.
We can write:
\begin{equation}\label{Gex}
\Gamma(R,h)_{ex} = 
\eta 
\Big( 
1 
+ {\mathbf  a }    {\mathbf \nabla} h 
+ ({ b_x }\frac{\partial^2}{\partial x^2}
+{ b_y }\frac{\partial^2}{\partial y^2} ) h  
\Big) \;\;\; .
\end{equation}
Here $\eta$ is the sputtering flux; ${\mathbf a}=(a_x,a_y)$ is a vectorial parameter associated with the flux-direction-dependent  erosion and $b_x$, $b_y$ are associated with the curvature-dependent sputtering erosion.
All these term are implicitly dependent on the orientation of the incoming ion flux, this in analogy with \cite{Bra88}.

A noise term $\xi(x,y,t)$ can be eventually added to Eq.\ref{Gex}.
This term might mimic the fact that the incoming flux is made -after all- of discrete particles and that in experiments certain amounts of noise and randomness are unavoidable.

\subsection{Adsorption}
The rate of adsorption of mobile atoms into immobile solid positions must be dependent on the quantity of mobile atoms in a given spatial position.
Similarly to the excavation process, the adsorption is also dependent on the local curvature and orientation. 
We can write:
\begin{equation}\label{Gad}
\Gamma(R,h)_{ad} =  
R 
\Big( 
\gamma 
+ {\mathbf  c }    {\mathbf \nabla} h 
+ ({ d_x }\frac{\partial^2}{\partial x^2}
+{ d_y }\frac{\partial^2}{\partial y^2} ) h
\Big) \;\;\;,
\end{equation}
where the parameter $\gamma$  is the recombination rate and ${\mathbf c}=(c_x,c_y)$ and ${ d_x }$, $d_y$  are associated to the different probabilities of recombination in relation with the local orientation and shape of the surface. 

Note that Eqs. \ref{Gex}, \ref{Gad} contain the same terms \cite{Laplace} as the ones proposed in the literature for the formation of aeolian dunes in the so-called hydrodynamical model \cite{Bouch95,Terz98,Val99,Csa00,Mak00}.
Indeed, in deserts, sand grains are lifted from the sand-bed and readsorbed into it with a probability which is dependent on the local shape and orientation of the dunes.
Eqs. \ref{Gex}, \ref{Gad} represent the simplest analytical expressions which formally take into account these shape and orientation dependences.
In the quest for simple explanations, such equations are therefore rather universal.

\subsection{Mobility}
Mobile atoms will move on the surface, and the quantity ${\mathbf J}(R,h) $ in Eq.\ref{E1} is the `current' of these atoms. 
In surface growth, there are two main mechanisms that are commonly indicated as responsible for the surface mobility of atoms \cite{Bar95}.
The first is a current, driven by the variations of the local chemical potential, which tends to smoothen the surface asperity moving atoms from hills to valleys. The second is a current induced by the Erlich-Schwoebel barrier which -on the contrary- moves atoms uphill.
In addiction to these main mechanisms we might also have to take into account a drift velocity and a random thermal diffusion, obtaining:
\begin{equation} \label{J1}
{\mathbf J}(R,h)  = 
{\mathcal K} R {\mathbf \nabla}({\mathbf \nabla}^2 h)
+ s R  \frac{ {\mathbf \nabla} h }{1+ ( \alpha_d {\mathbf \nabla} h)^2}
+
{\mathbf v } R 
- {\mathcal D } {\mathbf \nabla } R 
  \;\;\;.
\end{equation}
In this equation, the first term describes a deterministic diffusion driven by the variations of the chemical potential which depends on the local shape of the surface; the second term is associated with the uphill current due to the Erlich-Schwoebel barrier and $\alpha_d$ is a constant associated with the characteristic length.
The quantity ${\mathbf v }=(v_x,v_y)$ is a drift velocity of the mobile atoms on the surface, whereas ${\mathcal D }$ is the dispersion constant associated with the random thermal motion.
(The coefficient ${\mathcal D }$ is related with a non-deterministic diffusion mechanism and it could mimic the evaporation-condensation effects.)

Note that Eq.\ref{J1} is substantially different from the one proposed in the literature to describe ripples in granular media \cite{Bouch95,Terz98,Val99,Csa92,Csa93}. 
Here the current is supposed to be dependent on the local shape and orientation of the surface (the $ h({\mathbf r },t)$ profile). 
The equations describing sandy deserts can be retrieved from Eq.\ref{J1} by imposing ${\mathcal K}=0$ and $s=0$, but -on the contrary- in surface growth these two parameters are the leading terms of the equation and play the role of control parameters in the dynamics of ripple formation.
Nonetheless, these terms describe a rather simple dependence of the dynamic of particles on a surface on the geometrical shape of the surface itself.
Again, in our seek for universality, we expect that similar terms can be profitably introduced in the context of aeolian sand ripples in order to describe specific phenomena (associated, for instance, with packing properties \cite{Aste00} or granular flow \cite{Komatsu01}) which relate the current of grains with the dune-shapes.

It should be noted that the factors ${\mathbf a}$, ${\mathbf c}$ and ${\mathbf v}$ in Eqs.\ref{Gex},\ref{Gad} and \ref{J1} are \emph{vectors} (i.e. they have two distinct components in the $(x,y)$ directions).
Indeed, off-normal sputtering introduce anisotropy, and crystal surfaces are in general anisotropic. Therefore one must take into account the dependence of the parameters on the relative orientation of both the crystal-surface and the sputtering direction.
In Eq.\ref{J1} the deterministic diffusion parameter ${\mathcal K}$ is assumed isotropic.
On the other hand, on crystalline surfaces anisotropies are expected and sometimes might play an important role \cite{Park99}.
The extension of the results presented in this paper to the case of anisotropic ${\mathcal K}$ is straightforward.

%%%%%%%%%%%%%%%%%%%%%%%%%%%%%%%%%%%%%%%%%%%%%%%%%%%%%%%%%%%%%%%%%%%
%%%%%%%%%%%%%%%%%%%%%%%%%%%%%%%%%%%%%%%%%%%%%%%%%%%%%%%%%%%%%%%%%%%
\begin{figure}
%\hspace{+2cm}
\begin{center}
\includegraphics[width=0.75\textwidth]{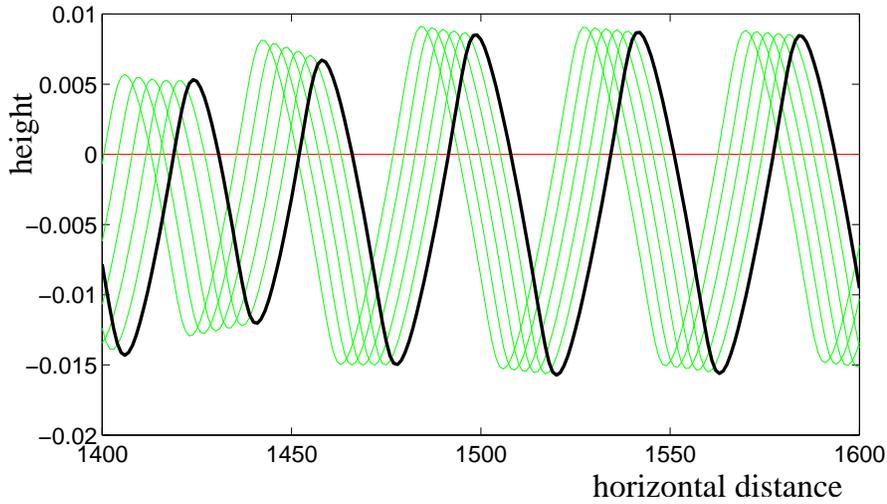}
\end{center}
\caption{\label{f.1} %\footnotesize
Numerical solutions of Eq.\ref{E1} at various times \cite{time} indicate that under the action of ion sputtering the surface develops an instability which leads to the formation of ripples with a well defined characteristic wavelength.
In the figure the black-tick line is the final surface-profile, whereas the thinner gray lines (green online) are some profiles at previous times.
See Appendix \ref{A.sim} for details.}
\end{figure}
%%%%%%%%%%%%%%%%%%%%%%%%%%%%%%%%%%%%%%%%%%%%%%%%%%%%%%%%%%%%%%%%%%%
%%%%%%%%%%%%%%%%%%%%%%%%%%%%%%%%%%%%%%%%%%%%%%%%%%%%%%%%%%%%%%%%%%%

%%%%%%%%%%%%%%%%%%%%%%%%%%%%%%%%%%%%%%%%%%%%%%%%%%%%%%%%%%%%%%%%%%%
%%%%%%%%%%%%%%%%%%%%%%%%%%%%%%%%%%%%%%%%%%%%%%%%%%%%%%%%%%%%%%%%%%%
\begin{figure}
%\hspace{+2cm}
\begin{center}
\includegraphics[width=0.75\textwidth]{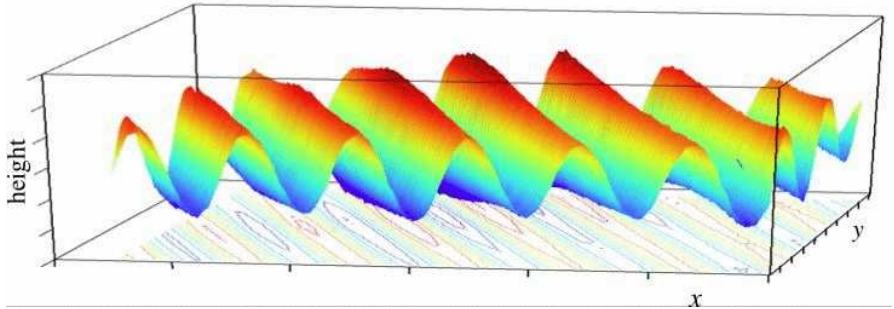}
\end{center}
\caption{\label{f.2D} %\footnotesize
A numerical solution of Eq.\ref{E1} in two dimensions.
See Appendix \ref{A.sim} for details.}
\end{figure}
%%%%%%%%%%%%%%%%%%%%%%%%%%%%%%%%%%%%%%%%%%%%%%%%%%%%%%%%%%%%%%%%%%%
%%%%%%%%%%%%%%%%%%%%%%%%%%%%%%%%%%%%%%%%%%%%%%%%%%%%%%%%%%%%%%%%%%%

\section{Dispersion Relation}

A trivial solution of Eq.\ref{E1} can be written for a completely flat surface: $h({\mathbf r },t)=h_0(t)$ and $R({\mathbf r },t)=R_0$.
In this case, we obtain $R_0 = (1-\phi) \eta/ \gamma$ and $h_0(t)=-\phi \eta t + const.$.
This describes a surface that rests flat and it is eroded with a speed equal to $\phi \eta$.
But this behavior is only hypothetical since -in general- the dynamics of the surface-profile presents instabilities against spontaneous roughening and therefore its evolution is more complex.
For instance, numerical solutions of Eq.\ref{E1}, are shown in Fig.\ref{f.1} (for the 1-dimensional case) and in Figs.\ref{f.2D}, \ref{f.rotation}, \ref{f.shape} (for the two-dimensional case).
We observe that, in a certain range of the parameters, the surface is unstable and periodic ripples or other instabilities are formed spontaneously.

\subsection{Stability analysis}

In order to infer indications about the amplification or the smoothing of small perturbations and to deduce an analytical expression for the ripples wave-length  at their beginning, we performe a stability analysis on Eq.\ref{E1}.
For this purpose we assume that the surface-profile is made by the combination of a flat term plus a rough part: 
\begin{eqnarray}\label{R1h1}
R({\mathbf r },t) &=& R_0 + R_1({\mathbf r },t) \nonumber \\
h({\mathbf r },t) &=& h_0(t) + h_1({\mathbf r },t) \;\;\;,
\end{eqnarray}
with $ R_1({\mathbf r },t) = \hat R_1 \exp(i \omega t + i {\mathbf k } {\mathbf r })$ and $ h_1({\mathbf r },t) = \hat h_1 \exp(i \omega t + i {\mathbf k } {\mathbf r })$. 
We substitute these quantities into Eq.\ref{E1} and linearize the equation by taking only the first order in $R_1$ and $h_1$.
A Fourier analysis (see Appendix \ref{A1}) shows that such a linearized equation admits solutions when the frequencies $\omega$ and the wave vectors ${\mathbf k}$ satisfy:
\begin{eqnarray} \label{Deter}
& &\left[ 
i \omega + 
\gamma 
+ i {\mathbf k } {\mathbf v } 
+  k^2 {\mathcal D } 
\right] \cdot
\nonumber \\
& &\left\{
i \omega 
+  i {\mathbf k } \left[ {\mathbf v }_1 - (1-\phi) {\mathbf v }_2 \right]
- k^2 \left[ {\mathcal D }_1 -(1-\phi) {\mathcal D }_2 \right]
\right\}-
\nonumber \\
& &
\gamma
(1-\phi)
\!
\left[
\!
i {\mathbf k } ({\mathbf v }_1 
\! - \!{\mathbf v }_2) 
\! - \! k^2  \left({\mathcal D }_1 
\! - \! {\mathcal D }_2 - s_1 \right)
\! - \! k^4 {\mathcal K }_1
\right] = 0
\;  ;
\nonumber \\
\end{eqnarray}
where, to simplify the equations, we have introduced the following notation: 
\begin{center}
\begin{tabular}{lll}
${\mathbf v}_1 = \eta {\mathbf a}$ 
%\hspace{1.cm} &  ${\mathbf v}_2 = \eta {\mathbf c}/{\gamma}$ \\
$s_1 = \eta s/{\gamma}$
& ${\mathcal K }_1 = \eta {\mathcal K }/\gamma$ \\
${\mathcal D }_1 = \eta (b_x \cos(\alpha)^2 + b_y \sin(\alpha)^2)$ \\
${\mathcal D }_2 = \frac{\eta}{{\gamma} } (d_x \cos(\alpha)^2 + d_y \sin(\alpha)^2)  \;\;\;.$\\
\end{tabular}
\end{center}
Where, $\alpha$ is the azimuthal angle corresponding the given direction of $\mathbf k$ in the $(x,y)$ plane (therefore $ k_x = |{\mathbf k}|\cos(\alpha)$, 
$ k_y = |{\mathbf k}|\sin(\alpha)$).
Equation \ref{Deter} establishes a \emph{dispersion relation} $\omega({\mathbf k})$ that is a complex function with two branches corresponding to the solutions of the quadratic Equation \ref{Deter}.

The dispersion relation Eq.\ref{Deter} reduces to the one from the linear analysis in the Bradley-Harper theory \cite{Bra88} when the coefficients $\mathcal D$, $\mathbf v$, $s$, $a_y$, $\gamma$, $\mathbf c$, $d_x$, $d_y$, $s_1$, ${\mathcal D}_2$ and ${\mathbf v}_2$ are set equal to zero.

\section{Surface Instabilities}

\begin{figure}
%\hspace{+2cm}
\begin{center}
\includegraphics[width=0.75\textwidth]{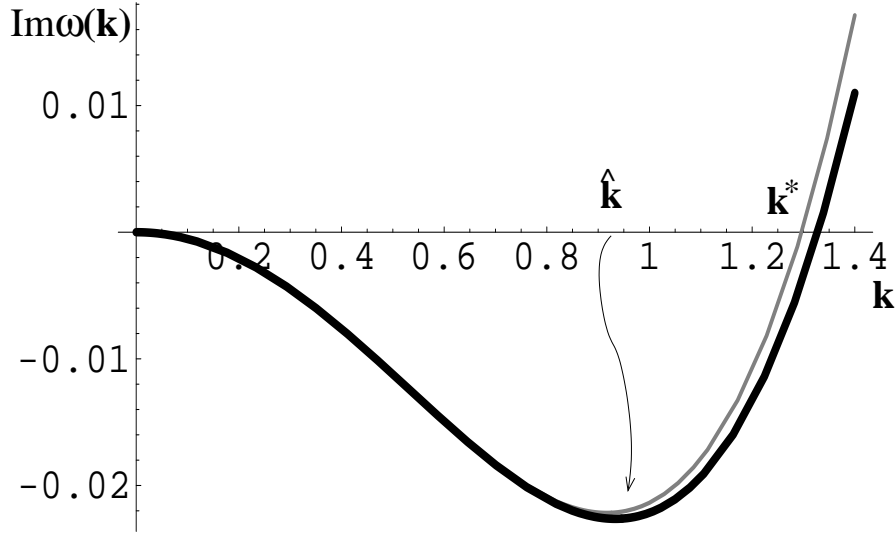}
\end{center}
\caption{\label{f.2} %\footnotesize
The imaginary part of the dispersion relation $Im(\omega)$ can assume negative values which are associated with the surface instability (arbitrary units).
The amplitude of modes with wavelengths $\lambda > 2\pi/k^*$ will grow exponentially fast.
The tick line is the imaginary part of the analytical solution of Eq.\ref{Deter}, whereas the thinny-gray line is the approximated expression (at the fourth order in $k$) obtained for small ion flux ($\eta$ small). 
(To draw this figure we set: $\gamma = 8$, $\phi = 0.5$, $v = 0.4$, $\mathcal D = 0.2$, ${\mathcal K}_1 = 0.06$, $s_1 = 0.09$, $v_1=0.01$, $v_2=0.001$, ${\mathcal D}_1 = 0.015$ and ${\mathcal D}_2 = 0.023$.)
}
\end{figure}

The kinetic growth of the surface instability is related to the immaginary part of $\omega({\mathbf k})$.
Indeed, $Im(\omega({\mathbf k}))$ corresponds to modes with amplitudes that change exponentially fast in time, and negative values correspond to unstable modes that increase with the time. 
We can therefore study $Im(\omega)$ from the solution of Eq.\ref{Deter} and search for the region of $\mathbf k$ in which $Im(\omega)$ is negative.
The most unstable mode is the one that grows faster and it corresponds to the value of ${\mathbf k}$ at which $Im(\omega)$ reaches its most negative value (see Figs.\ref{f.2} and \ref{f.2Dinstab}).

%%%%%%%%%%%%%%%%%%%%%%%%%%%%%%%%%%%%%%%%%%%%%%%%%%%%%%%%%%%%%%%%%%%
%%%%%%%%%%%%%%%%%%%%%%%%%%%%%%%%%%%%%%%%%%%%%%%%%%%%%%%%%%%%%%%%%%%
%%%%%%%%%%%%%%%%%%%%%%%%%%%%%%%%%%%%%%%%%%%%%%%%%%%%%%%%%%%%%%%%%%%
%%%%%%%%%%%%%%%%%%%%%%%%%%%%%%%%%%%%%%%%%%%%%%%%%%%%%%%%%%%%%%%%%%%

%%%%%%%%%%%%%%%%%%%%%%%%%%%%%%%%%%%%%%%%%%%%%%%%%%%%%%%%%%%%%%%%%%%
%%%%%%%%%%%%%%%%%%%%%%%%%%%%%%%%%%%%%%%%%%%%%%%%%%%%%%%%%%%%%%%%%%%
\begin{figure}
%\hspace{+2cm}
\begin{center}
\includegraphics[width=0.75\textwidth]{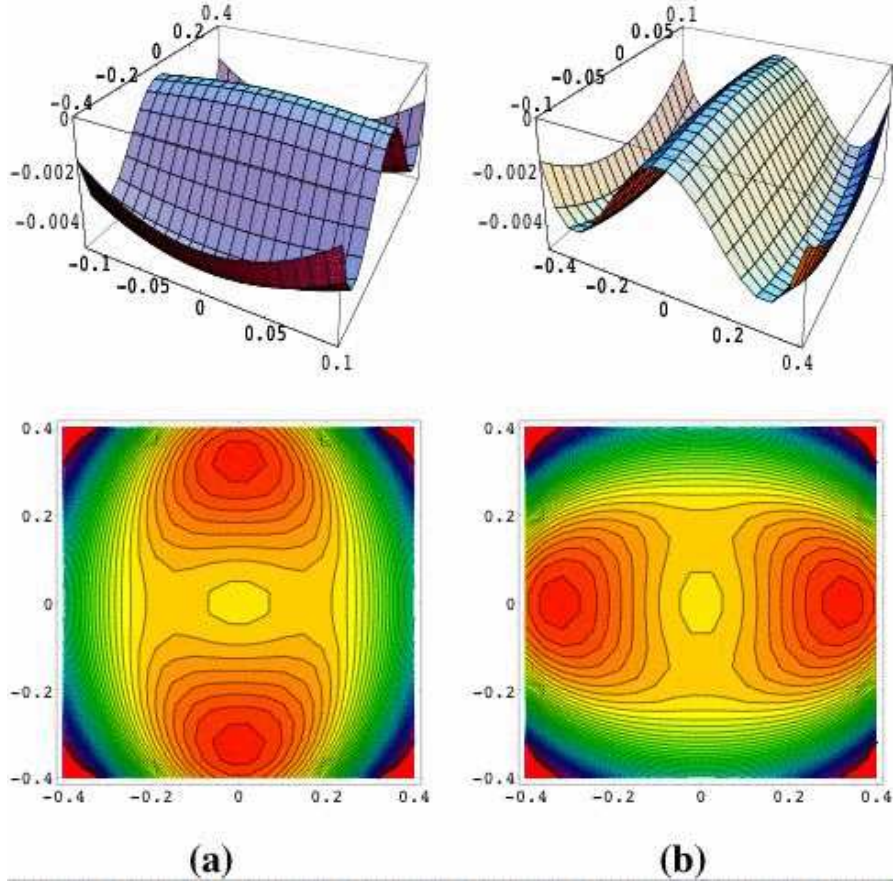}
\end{center}
\caption{\label{f.anis_2D} %\footnotesize
Surface-plot and contour lines plot for the imaginary part of the dispersion relation $Im(\omega)$ from the analytical solution of Eq.\ref{Deter} in the two-dimesional case.
Figures (a) and (b) show the occurrence of a 90 degrees rotation in the orientation of the instability mode associated with the relative variations of the coefficients $b_x$ and $b_y$.
To obtain the plots on the left side we set: $\mathbf v$, ${\mathbf v}_1$, ${\mathbf v}_2$ and $s$ = 0, $\gamma = 0.05$, $\eta = 1$, ${\mathcal D} = 0.1$, $b_x = 0.25$, $b_y = 0.5$, $d_x = 0.025$, $d_y = 0.025$, ${\mathcal K}_1 = 0.833$, $\phi = 0.2$. The right plots uses the same set of parameters with $b_x$ and $b_y$ exchanged.
}
\end{figure}
%%%%%%%%%%%%%%%%%%%%%%%%%%%%%%%%%%%%%%%%%%%%%%%%%%%%%%%%%%%%%%%%%%%
%%%%%%%%%%%%%%%%%%%%%%%%%%%%%%%%%%%%%%%%%%%%%%%%%%%%%%%%%%%%%%%%%%%

The solution of Eq.\ref{Deter} for $Im(\omega)$, is 
\begin{eqnarray}\label{solut}
2 Im(\omega)_\pm 
& = &
\gamma +
 \Big[{\mathcal D } - {\mathcal D }_1 +(1-\phi) {\mathcal D }_2 \Big]k^2
\nonumber \\
& \pm &
\sqrt{\frac{\Delta_1 + (\Delta_1^2 + 4 \Delta_2^2)^{1/2}}{2}}
\end{eqnarray}
where we have
\begin{eqnarray}\label{delta1}
\Delta_1 &=& 
\gamma^2 - \Big\{ \Big[ {\mathbf v} - {\mathbf v}_1 +(1-\phi) {\mathbf v}_2\Big] {\mathbf k} \Big\}^2
\nonumber \\
&+&
2 \gamma \Big[ {\mathcal D } - (1-2\phi){\mathcal D }_1 + (1-\phi) {\mathcal D }_2 + 2 (1-\phi)s_1\Big] k^2
\nonumber \\
&+& 
\Big\{
	\Big[({\mathcal D } + {\mathcal D }_1 - (1-\phi){\mathcal D }_2\Big]^2 
	- 4 \gamma (1-\phi) {\mathcal K }_1
\Big\} k^4
\nonumber \\
\end{eqnarray}
and 
\begin{eqnarray}\label{delta12}
\Delta_2 &=& 
\gamma \Big[{\mathbf v}+ (1-2\phi){\mathbf v}_1-(1-\phi){\mathbf v}_2\Big] {\mathbf k }
\nonumber \\
&+&
\Big[{\mathcal D } + {\mathcal D }_1- (1-\phi){\mathcal D }_2\Big] 
\Big[{\mathbf v}-{\mathbf v}_1+(1-\phi){\mathbf v}_2 \Big]{\mathbf k }^3 \;\;.
\nonumber \\
\end{eqnarray}

Let us first observe that in absence of sputtering (i.e. when $\eta = 0$ and therefore, ${\mathbf v }_1=0$, ${\mathbf v }_2=0$, 
${\mathcal D }_1=0$, ${\mathcal D }_2=0$,
$s_1=0$, ${\mathcal K }_1=0$) the solutions of Eq.\ref{Deter} are $\omega({\mathbf k}) =0$ and $\omega({\mathbf k}) = - {\mathbf k } {\mathbf v } + i (\gamma + k^2 {\mathcal D })$.
In this case, the imaginary part of $\omega({\mathbf k})$ is non-negative, therefore we -correctly- expect no spontaneous corrugation of the surface.
On the contrary, when the sputtering is active ($\eta \not= 0$), the immaginary part of $\omega({\mathbf k})$ can assume negative values.
This is shown in Fig.\ref{f.2} where a plot of $Im(\omega)_-$ is reported (along a given direction of the vector ${\mathbf k}$).
As one can see in Fig.\ref{f.2}, typically for a given direction of $\mathbf k$ one of the two branches $Im(\omega)_\pm$ takes negative values for $|{\mathbf k}|$ between 0 and a critical value $k^*$ at which it passes the zero.
The two-dimensional plot are given in Figs. \ref{f.anis_2D} and \ref{f.2Dinstab}.
Other cases are shown in Fig.\ref{f.instb} and discussed in Appendix \ref{A.Inst} together with the orientation-dependence.
The critical point $k^*$ (a contour in 2-dimensions, see Figs. \ref{f.anis_2D} and \ref{f.2Dinstab}) , fixes the minimal unstable wavelength.
We therefore expect to find unstable solutions associated with the formation and evolution of ripples with wavelengths $\lambda \ge \lambda^* = 2\pi/k^*$.

\section{Ripple wavelength}

Several analytical solutions of Eq.\ref{Deter} can be found in some special cases which are discussed in Section \ref{A2}.
But the study of the surface instabilities can be highly simplified if we consider the first order effects when the sputtering flux $\eta$ is small.

\subsection{Approximate equation}

In the case of small sputtering fluxes, the branch of $Im(\omega({\mathbf k}))$, with negative values can be approximated to:
\begin{equation}\label{dev}
Im(\omega)_-   
\simeq  
\frac{
P_1 k^6 + P_2 k^4 + P_3({\mathbf k})k^2 + P_4 k^2 + P_5({\mathbf k}) }{
{\mathcal D }^2 k^4 
+ 2 \gamma {\mathcal D } k^2 
+ ({\mathbf v  \mathbf k})^2
+\gamma^2  
}
\end{equation}
with
\begin{eqnarray}
&&P_1 = {\mathcal D } [ (1-\phi)\gamma {\mathcal K }_1 - {\mathcal D }({\mathcal D }_1 -(1-\phi) {\mathcal D }_2 )]
\nonumber \\
&&P_2 = (1-\phi) \gamma  [{\mathcal D }({\mathcal D }_2-s_1) + \gamma {\mathcal K }_1] - (1+\phi)\gamma {\mathcal D }{\mathcal D }_1  \nonumber \\
&&P_3({\mathbf k}) = - [{\mathcal D }_1 + (1-\phi) {\mathcal D}_2 ] ({\mathbf v}{\mathbf k})^2
\nonumber \\
&& P_4 = \gamma^2 [ (1-\phi) s_1 - \phi {\mathcal D }_1 ] 
\nonumber \\
&& P_5({\mathbf k}) = (1-\phi) \gamma ({\mathbf v}{\mathbf k}) [({\mathbf v}_1-{\mathbf v}_2){\mathbf k}] 
\end{eqnarray}

When $k = |{\mathbf k}|$ is sufficiently small ( $k \ll \gamma/\eta$ ), we can develop  Eq.\ref{dev} at the 4$^{th}$ order obtaining:
\begin{equation}\label{4ord}
Im(\omega)_-  \simeq  A k^4 - B k^2 \;\;,
\end{equation} 
with
\begin{eqnarray}\label{AB}
A &=& (1-\phi) \Big[
{\mathcal K }_1 
+ (s_1 + {\mathcal D }_2 -{\mathcal D }_1) \frac{ \gamma {\mathcal D } + v^2}{\gamma^2}+
\nonumber \\
& & v (v_1-v_2) \frac{ 2 \gamma {\mathcal D } + v^2}{\gamma^3}
\Big]
\nonumber \\
B &=& \phi {\mathcal D }_1 + (1-\phi) \Big[ s_1  + \frac{ v (v_1-v_2)}{\gamma} \Big]\;\;.
\end{eqnarray}
Here $v$, $v_1$  and $v_2$ are respectivelly the components of ${\mathbf v}$, ${\mathbf v}_1$  and ${\mathbf v}_2$ in the direction parallel to ${\mathbf k}$ (i.e. they are: $v = |{\mathbf v}|\cos(\widehat{{\mathbf v}{\mathbf k}})$; $v_1 = |{\mathbf v_1}|\cos(\widehat{{\mathbf v_1}{\mathbf k}})$ and $v_2 = |{\mathbf v_2}|\cos(\widehat{{\mathbf v_2}{\mathbf k}})$).

A comparison between this approximate solution and the exact one is given in Fig.\ref{f.2}.

Note that this approximate expression admits a negative minima, like the one shown in Fig.\ref{f.2}, only if the two terms $A$ and $B$ are both larger than zero.
Otherwise for $A >0$ and $B < 0$ the negative minimum disappears, whereas for $A<0$ the instability moves at $|{\mathbf k}|\to \infty$ (see Appendix \ref{A.Inst}, for a discussion of the exact case).

\subsection{Solutions}

The expected wavelength of the ripples is associated with the fastest growing mode, which corresponds to the value of ${\mathbf k}$ at which $Im(\omega)_-$ reaches its most negative point.
From Eq.\ref{4ord}, the minimum of $Im(\omega)$ is at 
\begin{equation}\label{khat}
\hat k = \sqrt{\frac{B}{2A}}\;\;\;.
\end{equation}
Therefore, at the beginning, the roughness will grow exponentially fast as $W \sim \exp(B^2 t/(4A))$ with associated ripple-wavelength at: \begin{equation}\label{wave}
\hat \lambda \sim 2 \pi \sqrt{\frac{2 A }{B }} \;\;\;\;.
\end{equation}

%%%%%%%%%%%%%%%%%%%%%%%%%%%%%%%%%%%%%%%%%%%%%%%%%%%%%%%%%%%%%%%%%%%
%%%%%%%%%%%%%%%%%%%%%%%%%%%%%%%%%%%%%%%%%%%%%%%%%%%%%%%%%%%%%%%%%%%
\begin{figure}
%\hspace{+2cm}
\begin{center}
\includegraphics[width=0.75\textwidth]{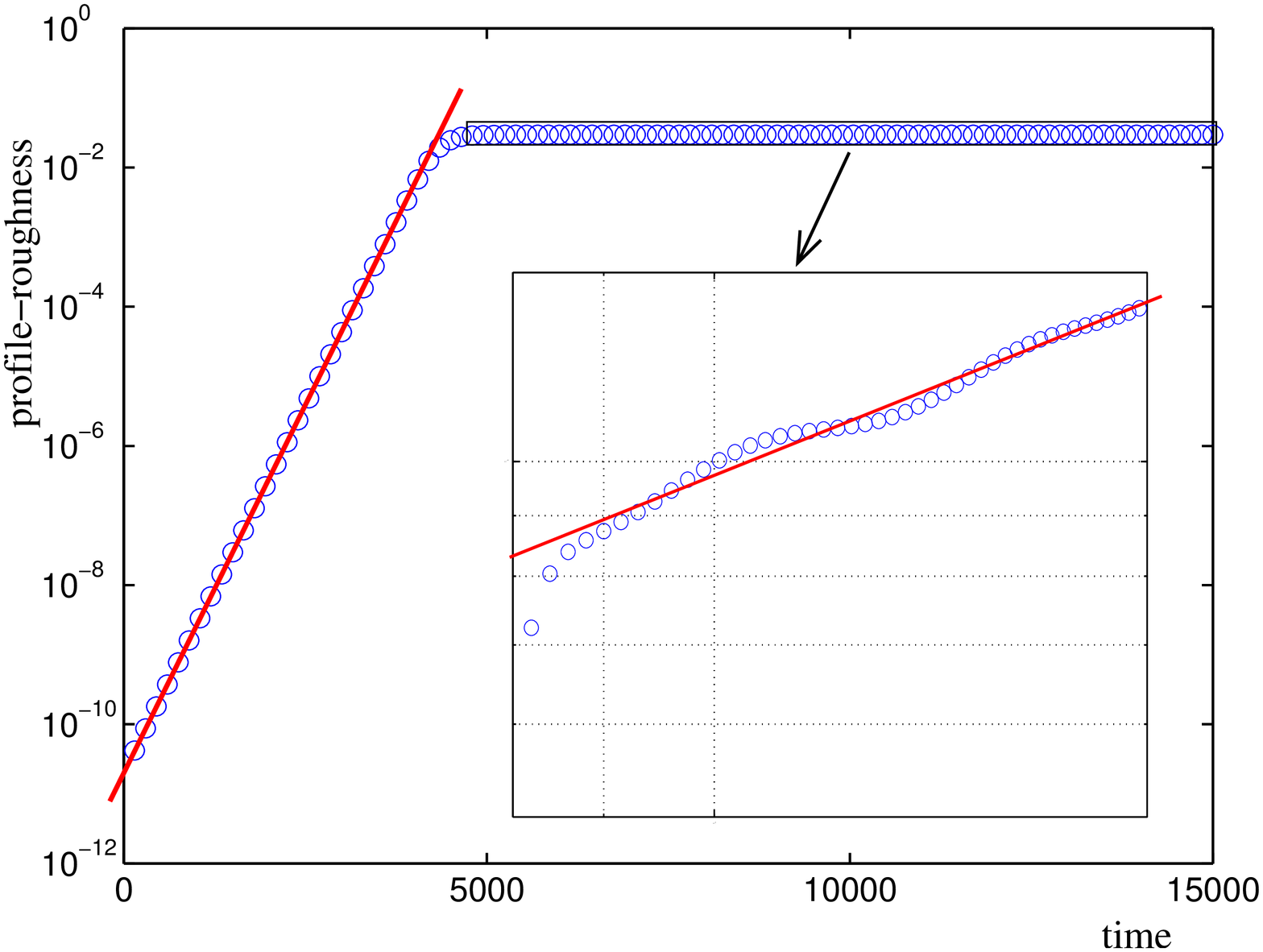}
\end{center}
\caption{\label{f.3} %\footnotesize
Evolution of the surface roughness (log-scale) v.s. time (linear scale) from numerical simulations (see Appendix \ref{A.sim}).
(Arbitrary units).
The insertion is the log-log plot of the last part of the evolution (highlighted with the rectangular box in the main plot).
When the Erlich-Schwoebel barrier is active the dynamical evolution of ripples can be described with an exponentially-fast growth at the beginning and then a `saturation' to a slower growth consistent with power-law (linear trend in log-log scale).
In this figure is also visible a slower behavior at the beginning of the growth-process. 
This transient regime, that often occurs in our numerical simulation, is also characterized by an exponential growth but with a longer characteristic time.
See Appendix \ref{A.sim} for details on the parameters used in the simulation.
}
\end{figure}
%%%%%%%%%%%%%%%%%%%%%%%%%%%%%%%%%%%%%%%%%%%%%%%%%%%%%%%%%%%%%%%%%%%
%%%%%%%%%%%%%%%%%%%%%%%%%%%%%%%%%%%%%%%%%%%%%%%%%%%%%%%%%%%%%%%%%%%

%%%%%%%%%%%%%%%%%%%%%%%%%%%%%%%%%%%%%%%%%%%%%%%%%%%%%%%%%%%%%%%%%%%
%%%%%%%%%%%%%%%%%%%%%%%%%%%%%%%%%%%%%%%%%%%%%%%%%%%%%%%%%%%%%%%%%%%
\begin{figure}
%\hspace{+2cm}
\begin{center}
\includegraphics[width=0.75\textwidth]{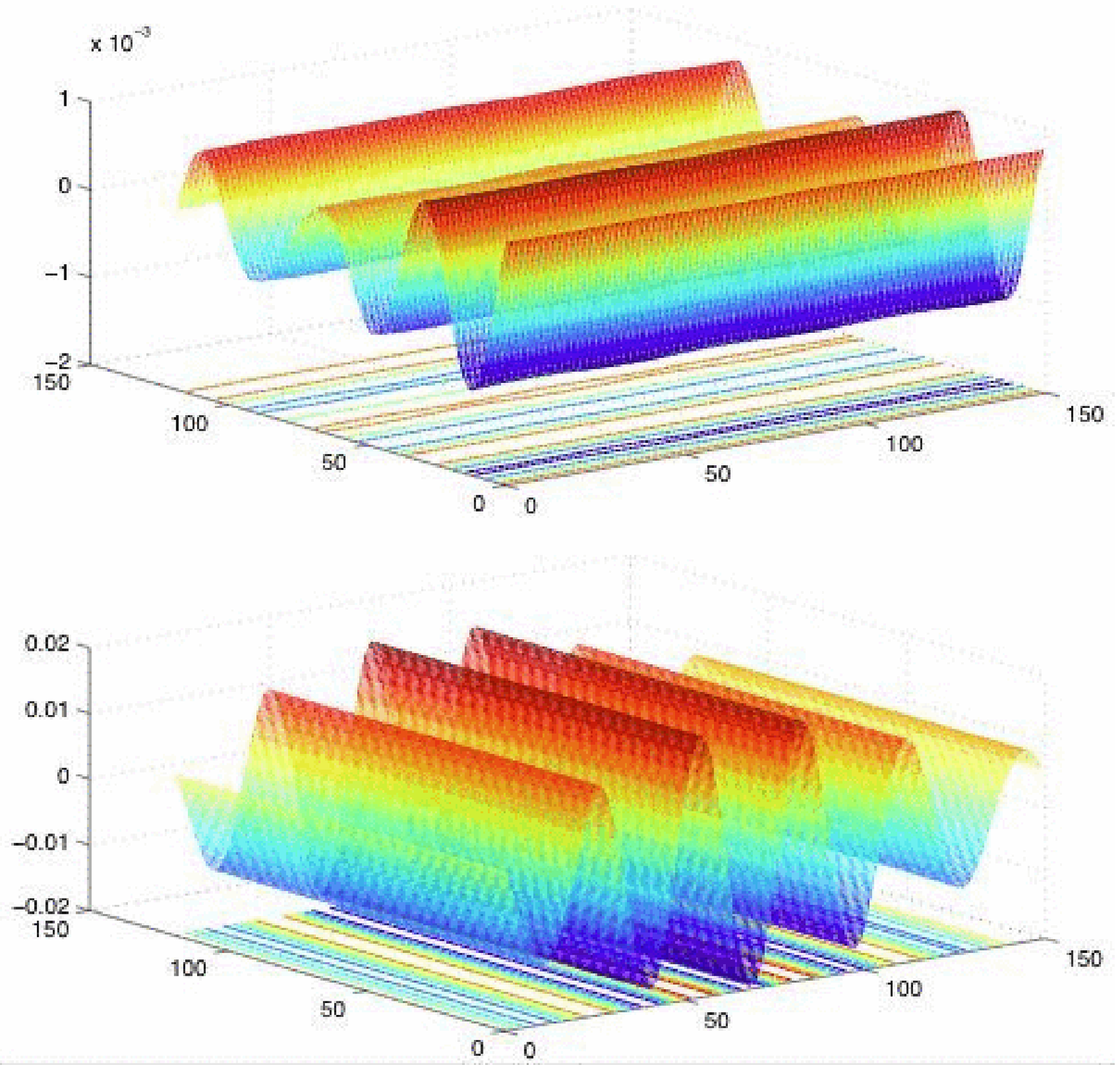}
\end{center}
\caption{\label{f.rotation} %\footnotesize
An example of ripples rotation associated with the relative variation of the two coefficients $b_x$ and $b_y$. 
See Appendix \ref{A.sim} for details. }
\end{figure}
%%%%%%%%%%%%%%%%%%%%%%%%%%%%%%%%%%%%%%%%%%%%%%%%%%%%%%%%%%%%%%%%%%%
%%%%%%%%%%%%%%%%%%%%%%%%%%%%%%%%%%%%%%%%%%%%%%%%%%%%%%%%%%%%%%%%%%%

\subsection{Orientation and Rotation}

The most unstable mode is selected by the position of the deepest minima in the $(k_x,k_y)$ plane. 
Equation \ref{khat} gives the radial position $|\mathbf{k}| = \hat k$ at which a section of $Im(\omega({\mathbf{k}}))$ along a given direction of $\mathbf{k}$ has a minimum. 
Such a minima corresponds to the value $Im(\omega({\mathbf{k}})) \sim - R^2/(4A)$ (Eq.\ref{4ord}).
The azimuthal position $\alpha$ of the absolute minimum is along the direction at which this minimum reaches the deepest point (see Figs. \ref{f.anis_2D} and \ref{f.2Dinstab}).
In Eq.\ref{solut} we have different sources of orientation-anisotropy.
First, the terms ${\mathcal D}_1$ and ${\mathcal D}_2$ which depend on the azimuthal angle.
Second, the quantities $v,v_1,v_2$ which also depend on such an angle.
Indeed, they are the components of the respective vectors ${\mathbf v, v_1, v_2}$ along the given direction of $\mathbf k$ ($v = |{\mathbf v}|\cos(\widehat{{\mathbf v}{\mathbf k}})$; $v_1 = |{\mathbf v_1}|\cos(\widehat{{\mathbf v_1}{\mathbf k}})$ and $v_2 = |{\mathbf v_2}|\cos(\widehat{{\mathbf v_2}{\mathbf k}})$). 
Consequently we expect the formation of ripples along preferential directions when $b_y$, $b_y$ and/or $d_x$, $d_y$ are anisotropic and/or when the parameters $\mathbf v$, $\mathbf v_1$, $\mathbf v_2$ are different from zero.
Indeed, in Figs. \ref{f.anis_2D} and \ref{f.2Dinstab} we show that the azimuthal position of the minimum can rotate by varying the relative weight of the coefficients $b_x$ and $b_y$.
Numerical solutions, reported in Fig.\ref{f.rotation}, indicate that ripples form along preferential directions and rotations of their orientations can be induced by introducing anisotropy in the $b_x$ and $b_y$ coefficients.
Other kinds of ripple-rotations can be induced by varying the $x$-, $y$-components of the vectors ${\mathbf v}$, ${\mathbf v_1}$, ${\mathbf v_2}$ (Fig.\ref{f.2Dinstab}). An example of ripple-rotation induced by the variations of these terms is provided in Fig.\ref{f.shape}.

The dependence of the ripple orientation from the incidence angle of the ion-sputtering can be easily explained by the associated dependence of the adsorption/excavation rates and the drift from the sputtering direction.
We also observe that in the isotropic case and when the non-scalar quantities $\mathbf v$, $\mathbf v_1$ and $\mathbf v_2$ are set to zero the surface instability tends to generate spontaneously pyramidal-like surface-structures (see Fig.\ref{f.shape}).
For a further discussion of the orientation-dependence of the instability modes see Appendix \ref{A.Inst}.

\subsection{ Some special cases}

Let us first observe that, when ${\mathcal K }_1 $, $s_1$ and $\phi$ are equal to zero, the ripple wavelength, given by Eq.\ref{wave}, coincides with the one found for sand dunes in deserts (see for instance \cite{Val99}). 
In our notation the `reptation length' is $l_0 = v/\gamma$, the `cut-off length' is $l_c = ({\mathcal D }_2-{\mathcal D }_1)/v$, whereas $v_1-v_2$ is the collective drift velocity of the dunes.
The approximations usually applied in this context \cite{Terz98,Val99}, imply: $l_c \gg \sqrt{{\mathcal D }/\gamma}$, and $\gamma l_c \gg v_1-v_2$,
giving, from Eq.\ref{wave}
\begin{equation}
\hat \lambda \sim 2 \pi \sqrt{ \frac{2 v l_0 l_c }{v_1-v_2} } \;\;\;\;.
\end{equation}

Let us now consider the dynamical evolution of a surface under ion sputtering and in particular the case when the effect of the Erlich-Schwoebel barrier is not present (as for semiconductors and glasses). 
In this case, $s=0, s_1=0$ and we also expect that the drift velocity $v$ and the dispersion constant ${\mathcal D }$ are equal to zero or infinitesimally small.
Indeed, here the current of mobile atoms on the surface is mainly induced by the differences in the chemical potential. 
Under these assumptions, from Eq.\ref{wave}, the wavelength of the most unstable ripple is:
\begin{equation}\label{lk}
\hat \lambda \sim 2 \pi \sqrt{ \frac{2 {\mathcal K }}{\nu}} \;\;\;\;,
\end{equation}
where we called $\nu = \frac{\gamma}{\eta} {\mathcal D}_1 \phi/(1-\phi) $, a quantity which plays the role of an effective surface tension.
Note that Eq. \ref{lk} is the same result as from the Bradley and Harper theory \cite{Bra88,Park99,Bar95,Cuer95,Gill01}.

When the Erlich-Schwoebel barriers are active ($s,s_1 \not= 0$), effects can be observed on the ripple-wavelength at their beginning, which becomes:
\begin{equation} \label{lks}
\hat \lambda \sim 2 \pi \sqrt{\frac{2 {\mathcal K }}{\nu + s }} \;\;\;\;.
\end{equation}

\section{Exact solutions} \label{A2}

Compact analytical expressions for the values of $\mathbf k$ at which $Im(\omega)=0$ (${\mathbf k}^*$) can be calculated from Eq.\ref{solut} in some special cases. 

In particular, when $\phi=0$, $s_1=0$, ${\mathcal K }_1=0$ and ${\mathcal D }=0$, we obtain
\begin{equation}\label{a0}
k^*
=
\sqrt{
\frac{
 \gamma ({v }_1 -{ v }_2)   }{ 
(v - { v }_1 + { v }_2)
 ({\mathcal D }_2 - {\mathcal D }_1) 
} 
}  
\;\;\;,
\end{equation}
where $v$, $v_1$ and $v_2$ are the components of ${\mathbf v}$, ${\mathbf v}_1$ and ${\mathbf v}_2$ in the direction of ${\mathbf k }^*$.
Whereas the value $k^*$ is the radial component of $\mathbf k$ along the closed contour in the $(k_x,k_y)$ plane at which the surface $Im(\omega({\mathbf k}))$ crosses the zero (see Figs. \ref{f.anis_2D} and \ref{f.2Dinstab}).

On the other hand when, ${\mathcal K }_1$, $s_1$, ${\mathcal D }_1$ and ${\mathcal D }_2$ are equal to zero, we find 
\begin{equation}\label{a3}
k^*
=
\sqrt{
\frac{
\gamma ({v } - \phi { v }_1)
}
{
\mathcal D ({v }_1 - { v } -(1-\phi) { v }_2) 
}} \;\;\;.
\end{equation}

The effect of the deterministic diffusion induced by the chemical potential can be studied from the solution  
\begin{equation}\label{a1}
k^*
=
\sqrt{
\frac{\phi \gamma {\mathcal D }_1}{ 
(1-\phi) \gamma {\mathcal K }_1  - {\mathcal D }({\mathcal D }_1 - (1-\phi) {\mathcal D }_2) } }  \;\;\;,
\end{equation}
which holds when ${\mathbf v }=0$, ${\mathbf v }_1=0$, ${\mathbf v }_2=0$, $s_1=0$ and ${\mathcal D }-{\mathcal D }_1 + (1-\phi) {\mathcal D }_2 >0$.

Whereas, when we set to zero ${\mathbf v }_1$, ${\mathbf v }_2$, ${\mathcal D }_1$ and ${\mathcal D }_2$, we find 
\begin{equation}\label{a2}
k^*
=
\sqrt{
\frac{ s_1 }{ 
{\mathcal K }_1  } }  \;\;\;,
\end{equation}
\noindent
which implies that the uphill current due to the Erlich-Schwoebel barrier can generate instability even when the shape-dependent erosion and recombination terms are inactive.

%%%%%%%%%%%%%%%%%%%%%%%%%%%%%%%%%%%%%%%%%%%%%%%%%%%%%%%%%%%%%%%%%%%
%%%%%%%%%%%%%%%%%%%%%%%%%%%%%%%%%%%%%%%%%%%%%%%%%%%%%%%%%%%%%%%%%%%
\begin{figure}
%\hspace{+2cm}
\begin{center}
\includegraphics[width=0.75\textwidth]{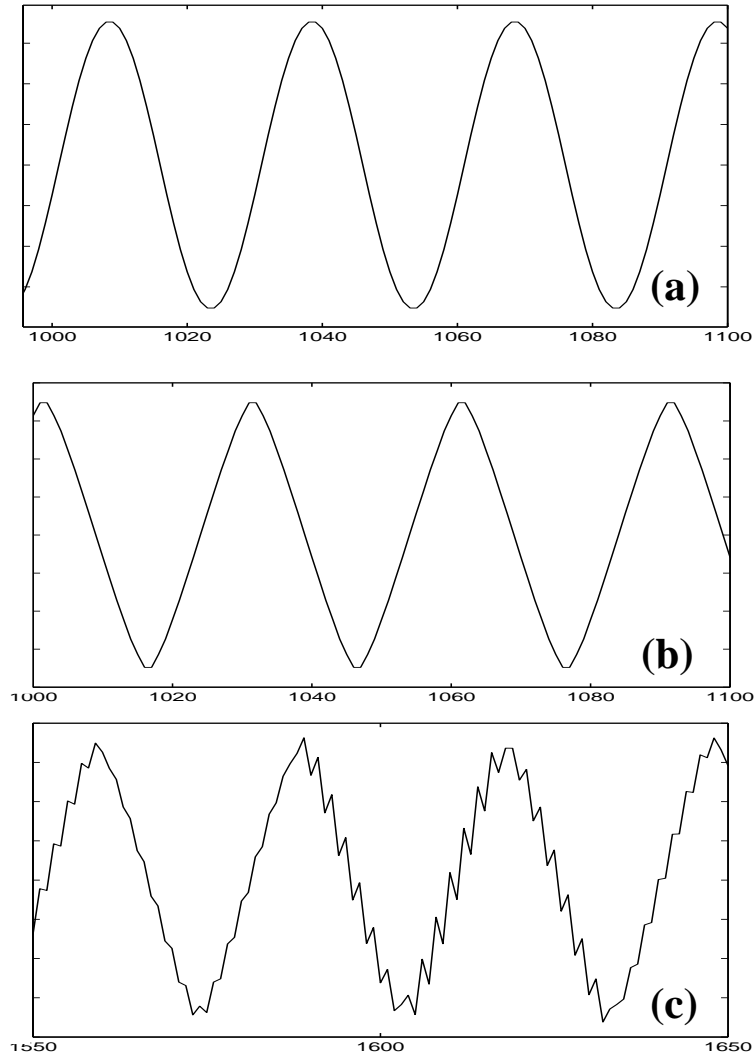}
\end{center}
\caption{\label{f.saturation} 
%\footnotesize
During the exponential growth the profile shows a typical sinusoidal shape (a). 
Whereas at the saturation/stabilization, the profile tends to eliminate the curvature creating more triangular-like shapes (b).
At this stage instabilities might take place, with short wave-length profile corrugations which nucleate on the triangular-like profile (c).
(See text.)}
\end{figure}
%%%%%%%%%%%%%%%%%%%%%%%%%%%%%%%%%%%%%%%%%%%%%%%%%%%%%%%%%%%%%%%%%%%
%%%%%%%%%%%%%%%%%%%%%%%%%%%%%%%%%%%%%%%%%%%%%%%%%%%%%%%%%%%%%%%%%%%
%%%%%%%%%%%%%%%%%%%%%%%%%%%%%%%%%%%%%%%%%%%%%%%%%%%%%%%%%%%%%%%%%%%
%%%%%%%%%%%%%%%%%%%%%%%%%%%%%%%%%%%%%%%%%%%%%%%%%%%%%%%%%%%%%%%%%%%
\begin{figure}
%\hspace{+2cm}
\begin{center}
\includegraphics[width=0.75\textwidth]{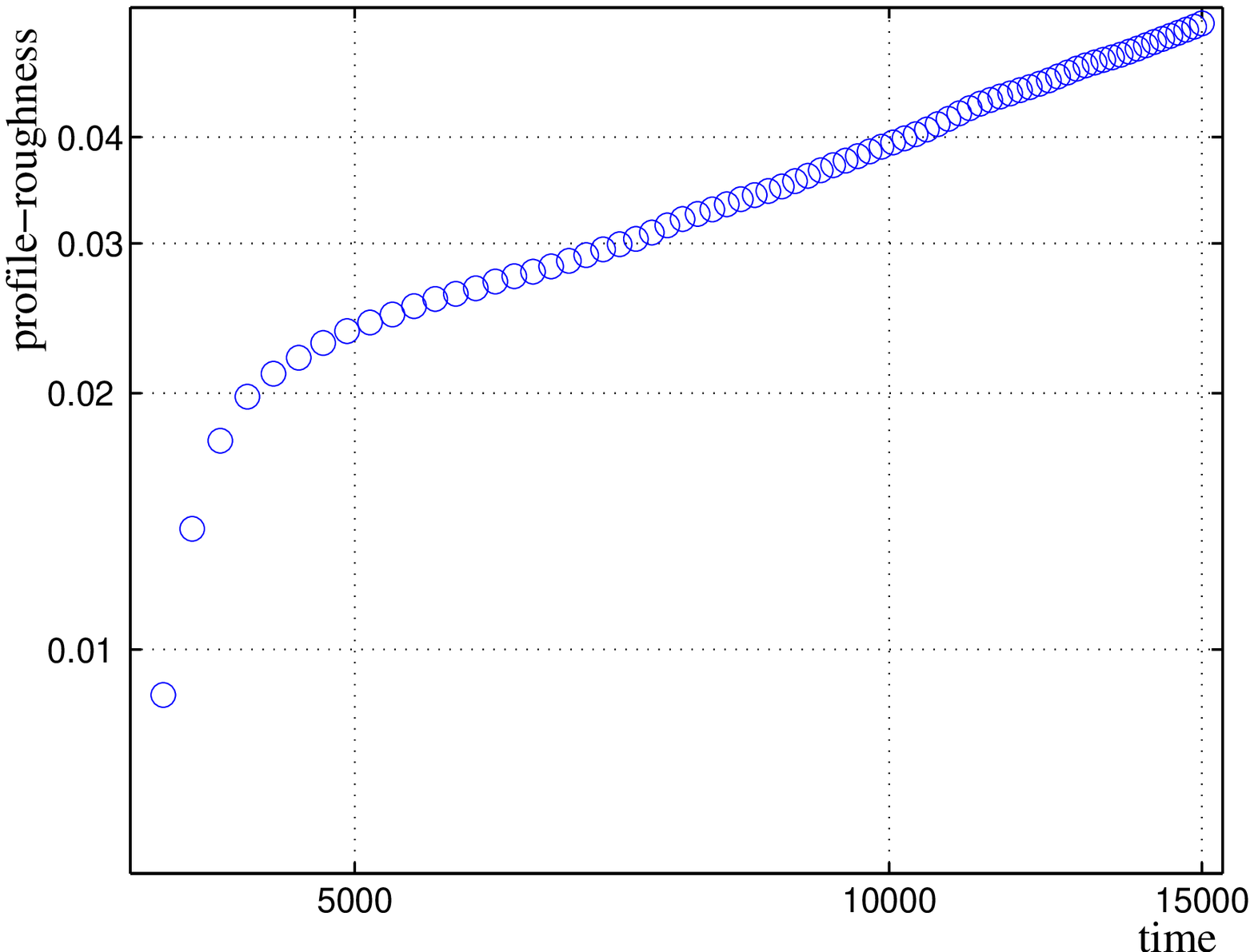}
\end{center}
\caption{\label{f.satNonUnif} 
%\footnotesize
When instabilities re- nucleate on the triangular-like profiles the growth is characterized by a power-law regime. 
(See text.)}
\end{figure}
%%%%%%%%%%%%%%%%%%%%%%%%%%%%%%%%%%%%%%%%%%%%%%%%%%%%%%%%%%%%%%%%%%%
%%%%%%%%%%%%%%%%%%%%%%%%%%%%%%%%%%%%%%%%%%%%%%%%%%%%%%%%%%%%%%%%%%%

\section{Exponential growth, stabilization, saturation and critical roughness}
\label{s.sat}

In metals, when the Erlich-Schwoebel barrier is active, there is an important non-linear contribution in the current of mobile atoms which becomes sizable when the roughness becomes sufficiently large and therefore $\left< (\alpha_d {\mathbf \nabla} h)^2 \right> \sim 1$ (see Eq.\ref{J1}) (the average is over the surface positions).
We observe -numerically- that this term changes the ripple's growth dynamics: from exponential to a slower growth consistent with a power-law.
Sometime a saturation to a constant value is observed.
This effect is shown in Figs.\ref{f.3} and \ref{f.satNonUnif}.
Numerically, all the computed exponents follow in a wide range between $0$ (saturation) to $1$.
We observe that this power-law regime is strongly affected by the presence or the absence of a noise term in Eq.\ref{Gex}.
In particular the saturation to a fixed profile with constant roughness has been observed only in absence of the noise term.
In the power-law-like regime the ripple-wavelength tends also to grow with time.
A theoretical justification of this power law regime and the evaluation of the exponent is under current investigation. 
From preliminary studies (described in Appendix \ref{A.stab}) it seems apparent that this phase of the growth is characterized by a strong non-linearity with a strong dependence of the ripple shapes (Fig.\ref{f.saturation}) and dynamics (Figs.\ref{f.3} and \ref{f.satNonUnif}) from small variations of the system parameters or the starting conditions.
These preliminary studies seem to indicate that this regime is governed by the re-nucleation of new instabilities over the rippled surface.
Experimentally, power law growth of the roughness and growth of characteristic wavelengths were observed in erosive sputtering \cite{Valbusa02}.
On the other hand, re-nucleation of instabilities has not been reported yet.

In semiconductors or glasses, when no Erlich-Schwoebel barrier is present, it is physically intuitive that the exponential growth of the surface roughness (which is a characteristic of the beginning of the surface instability) cannot continue indefinitely. 
Indeed, from the expression $ R({\mathbf r },t) = R_0 + \hat R_1 \exp(i \omega t + i {\mathbf k } {\mathbf r })$, which we used to derive Eq.\ref{wave}, we can immediately observe that when $\hat R_1 > R_0 = (1-\phi) \eta/ \gamma $, the amount of mobile atoms might become negative.
Since a negative amount of atoms is physically impossible, the process of exponential roughness-growth described above must necessarily finish around a critical roughness given by:
\begin{equation}\label{Wc}
W_c \sim (1-\phi) \frac{\eta}{\gamma} \;\;\;.
\end{equation} 
This behavior is confirmed by numerical solutions of Eq.\ref{E1} \cite{NumIntab} and it is expected to be observable in semiconductors and glasses after sufficiently long times.

We note that in the present section, these nonlinear effects are only sketched and this might lead to some incompleteness and trivialities. 
The systematic study of the nonlinear behavior is a worthwhile exercise which requires further careful investigations and it would be the subject of future publications.

\section{Conclusions}

We have shown that the same theoretical approach introduced to describe the formation of aeolian sand ripples can be conveniently applied to the study of the formation of periodic structures on surfaces under ion sputterning.
We have two physically different phenomena which both take place at a surface and involve two kind of particles.
In both cases these particle are excavated, adsorbed and displaced on the surface by different physical agents.
The equations used to describe the Excavation and Adsorption mechanisms (Eqs. \ref{Gex}, \ref{Gad}) are the simplest expressions that take into account a dependence of these phenomena to the surface shape. 
Indeed, they are composed by a constant term, plus a first derivative, plus a second derivative (i.e. height/orientation, slope, and curvature). 
Such terms must appear in any theory that addresses the problem of surface excavation and adsorption. 
On the other hand, the mobility term contains more specific terms like the deterministic diffusion or the Erlich-Schwoebel barrier term, which are specific of the ion sputtering surface physics.
But, also in this case, these terms describe a rather simple dependence of the particle mobility form the surface shape.
Similar terms might be profitably introduced in the description of sand dune dynamics.

We performed a linear stability analysis for pattern formation under sputtering erosion and we obtained general expressions for the ripples wavelength in term of the system parameters.
It has been shown that in some particular cases such solutions coincide with the ones already known in the literature for sand dunes and surface instability \cite{Bra88,KPZ,Park99,Terz98,Val99}.
We have discussed the effect of the Erlich-Schwoebel barriers and compared the result with numerical solutions.
We pointed out the Erlich-Schwoebel barrier can be responsible for a dramatic change in the system dynamics: from the exponential growth to a slower growth compatible with a power-law dynamics.
The occurrence of a critical roughness has been predicted.
The effect of ripple-renucleation in this regime has been also highlighted.
The dependence of the ripple-orientation and shapes in function of the system parameters has been discussed and the intriguing phenomenon of `ripple-rotation' has been accounted.

It should be noted that the main purpose of this paper is to point out a relevant example of universality: two processes which have completely different scales present a dynamical evolution which obeys to the same geometrical constraints and thus can be described by using the same phenomenological model.
On the other hand, we must observe that the class of solutions of Eq.\ref{E1} is rich and complex - even in the linear approximation.
Exhaustive, systematic studies of the classes of solutions of this equation and their dependence on the set of parameters will be the subject of future studies and publications.

\subsection*{Acknowledgments}
T. Aste thanks the Australian Partnership for Advanced Computing for the use of the (super-)Computing National Facility.
%We thank the referees for their careful scrutiny of our manuscript and their suggestions. 

\appendix

\section{Fourier transform of the linearized equation} \label{A1}

By substituting Eqs.\ref{R1h1}, \ref{J1}, \ref{Gad} and \ref{Gex} into Eq.\ref{E1} and by neglecting the second order terms (in $R_1$ and $h_1$), we obtain the following linearized equation:
\begin{eqnarray} \label{EqStab}
\frac{\partial h_1}{\partial t} 
&=&  
   \gamma R_1 
- \left[ {\mathbf v}_1 - (1-\phi){\mathbf v}_2 \right] {\mathbf \nabla} h_1 
- 
\nonumber \\
& &
\left[ {\mathcal D }_1 - (1-\phi){\mathcal D }_2 \right] {\mathbf \nabla}^2 h_1 
\nonumber \\
\frac{\partial R_1}{\partial t} 
&=& 
- \gamma R_1
- {\mathbf v }  {\mathbf \nabla } R_1 
+ {\mathcal D } {\mathbf \nabla }^2 R_1 +
\nonumber \\
&& (1-\phi)\Big[ ({\mathbf v}_1-{\mathbf v}_2)   {\mathbf \nabla}  h_1 +   
\nonumber \\
&& ({\mathcal D }_1 -{\mathcal D }_2 - s_1    ) {\mathbf \nabla}^2 h_1
- {\mathcal K }_1 {\mathbf \nabla}^4 h_1 \Big]
\;\;\;\;.
\end{eqnarray}

A Fourier analysis of Eq.\ref{EqStab} leads to
\begin{eqnarray} \label{EqFou}
& &
\gamma  \hat R_1 
-\Big\{
i \omega 
+  
i {\mathbf k } \left[ {\mathbf v }_1 - (1-\phi) {\mathbf v }_2 \right] -
\nonumber \\
& & 
k^2 \left[ {\mathcal D }_1 -(1-\phi) {\mathcal D }_2 \right]
\Big\} \hat h_1 = 0
\nonumber \\
& &
\left[ 
i \omega + 
\gamma 
+ i {\mathbf k } {\mathbf v } 
+  k^2 {\mathcal D } 
\right]  \hat R_1 
-
(1-\phi)
\Big[
  i {\mathbf k } ({\mathbf v }_1 -{\mathbf v }_2) 
-
\nonumber \\
& &
k^2  \left({\mathcal D }_1 - {\mathcal D }_2 - s_1 \right)
- k^4 {\mathcal K }_1
\Big] \hat h_1 = 0
\;\;\;,
\end{eqnarray}
with $\hat R_1$ and $\hat h_1  $ the Fourier components of $R_1$ and $h_1$ respectively.
This equation is a simple linear equation in two variables. 
It admits a non-trivial solution when the determinant of the coefficients is equal to zero.
This leads to Eq.\ref{Deter}.

%%%%%%%%%%%%%%%%%%%%%%%%%%%%%%%%%%%%%%%%%%%%%%%%%%%%%%%%%%%%%%%%%%%
%%%%%%%%%%%%%%%%%%%%%%%%%%%%%%%%%%%%%%%%%%%%%%%%%%%%%%%%%%%%%%%%%%%
\begin{figure}
%\hspace{+2cm}
\begin{center}
\includegraphics[width=0.75\textwidth]{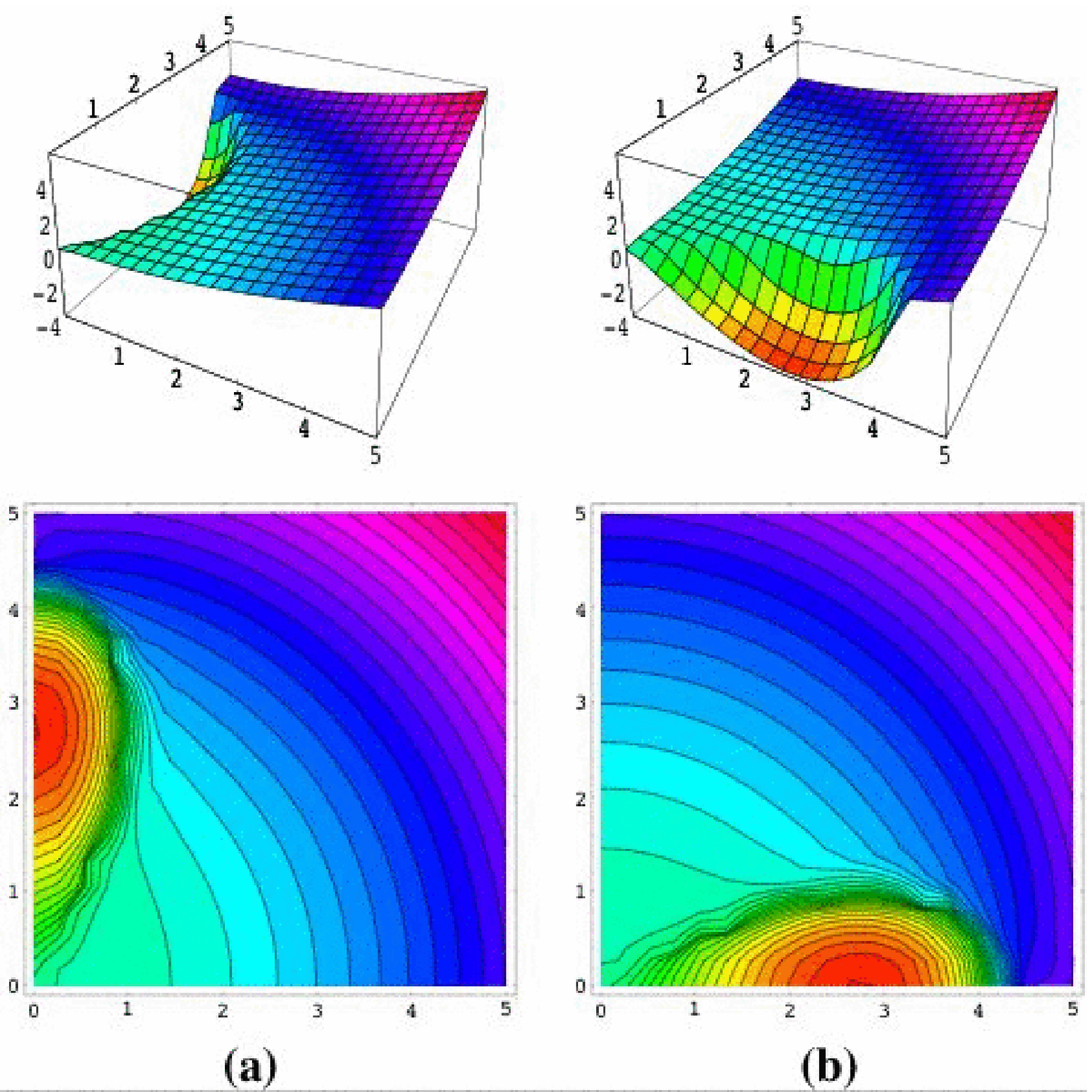}
\end{center}
\caption{\label{f.2Dinstab} %\footnotesize
Surface-plot and contour lines plot for the imaginary part of the dispersion relation $Im(\omega)$ from the analytical solution of Eq.\ref{Deter} in the two-dimesional case.
Figures (a) and (b) show the occurrence of a 90 degrees rotation in the orientation of the instability mode associated with the variations of the components in the $x,y$ plane of the coefficient $\mathbf v_1$.
To obtain the plots on the left side we set: $ v_x = 0.1$,$ v_y = 0.1$, $(v_1)_x=10$, $(v_1)_y=0.001$, ${\mathbf v}_2=0$, $s=7$, $\gamma = 1$, $\eta = 1$, ${\mathcal D} = 0.1$, , ${\mathcal D}_1 = 0.7$, ${\mathcal D}_2 = 1$, ${\mathcal K}_1 = 0.4$, $\phi = 0.2$. The right plots use the same parameters with exchanged components for ${\mathbf v}_1$.
}
\end{figure}
%%%%%%%%%%%%%%%%%%%%%%%%%%%%%%%%%%%%%%%%%%%%%%%%%%%%%%%%%%%%%%%%%%%
%%%%%%%%%%%%%%%%%%%%%%%%%%%%%%%%%%%%%%%%%%%%%%%%%%%%%%%%%%%%%%%%%%%
%%%%%%%%%%%%%%%%%%%%%%%%%%%%%%%%%%%%%%%%%%%%%%%%%%%%%%%%%%%%%%%%%%%
%%%%%%%%%%%%%%%%%%%%%%%%%%%%%%%%%%%%%%%%%%%%%%%%%%%%%%%%%%%%%%%%%%%
\begin{figure}
%\hspace{+2cm}
\begin{center}
\includegraphics[width=0.75\textwidth]{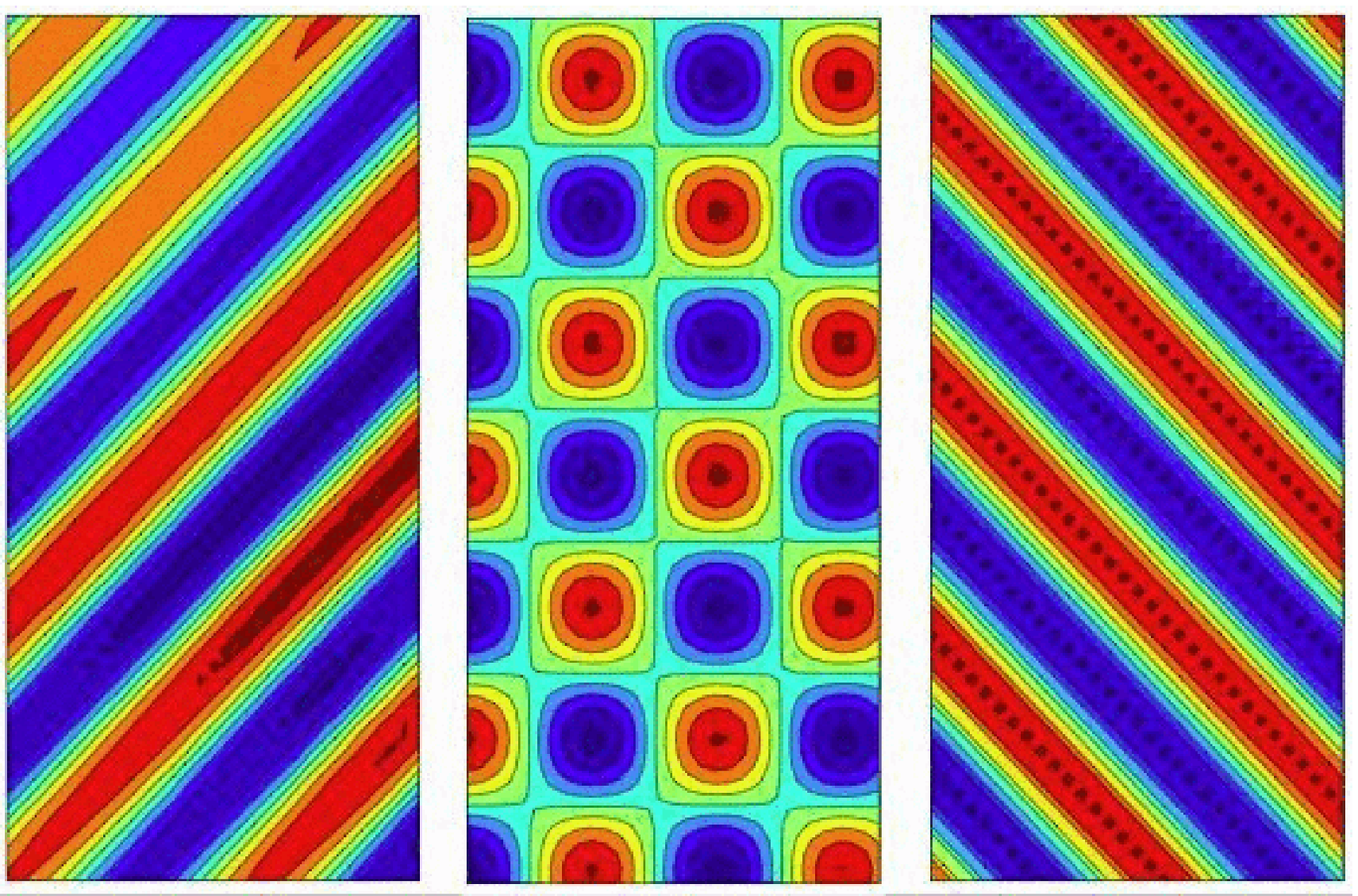}
\end{center}
\caption{\label{f.shape} %\footnotesize
An example of ripples rotation (left and right figures) associated with the variation of the components of $\mathbf v$, $\mathbf v_1$, $\mathbf v_2$ in the $(x,y)$ plane. 
Different profile instability (from modulated rippled-structure to pyramidal-like surface-structure) might take place when the parameters $\mathbf v$, $\mathbf v_1$ and $\mathbf v_2$ go to zero (central figure).
}
\end{figure}
%%%%%%%%%%%%%%%%%%%%%%%%%%%%%%%%%%%%%%%%%%%%%%%%%%%%%%%%%%%%%%%%%%%
%%%%%%%%%%%%%%%%%%%%%%%%%%%%%%%%%%%%%%%%%%%%%%%%%%%%%%%%%%%%%%%%%%%

\section{Orientation dependence of the instability} \label{A.Inst}

There are three distinct mechanisms that can lead to ripple-rotation.

The {\bf first} is associated with the anisotropies in the curvature-dependent erosion coefficients $b_x$, $b_y$ and in the curvature-dependent adsorption coefficients $d_x$, $d_y$. 
These anisotropies are induced both by the sputtering orientation angle and by the crystal orientation (in non-amorphous surfaces).
Example of the instability rotation and ripple re-orientation are given in Figs.\ref{f.anis_2D} and \ref{f.rotation}.

\begin{figure}
%\hspace{+2cm}
\begin{center}
\includegraphics[width=0.75\textwidth]{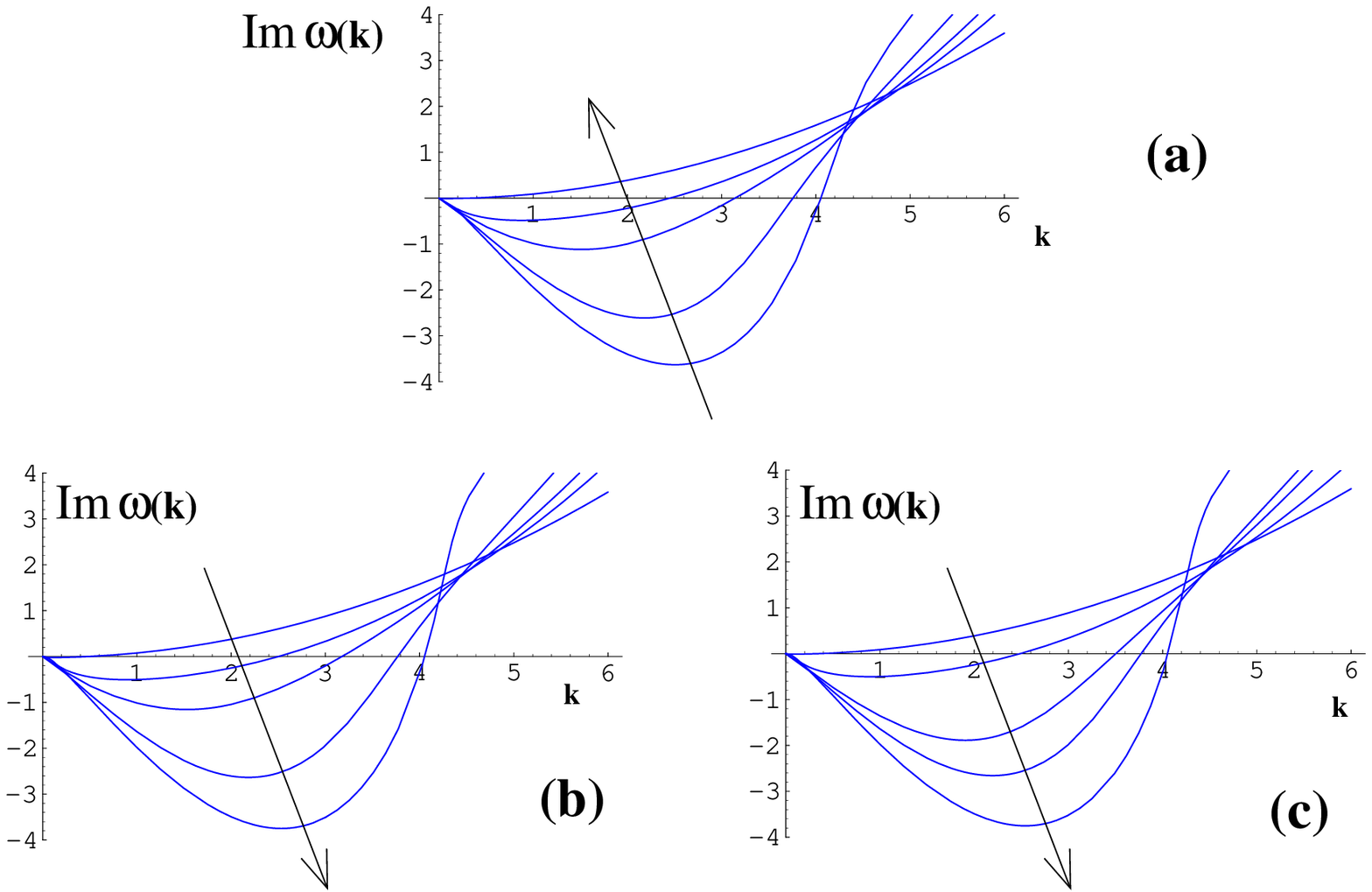}
\end{center}
\caption{\label{f.instb} 
%\footnotesize
The imaginary part of the dispersion relation $Im(\omega)$ v.s. $|\mathbf{k}|$ along a given azimuthal direction for various values of the components of $\mathbf{v,v_1,v_2}$.
Instabilities become deeper or rise and even disappear depending on the weights of the components of $\mathbf{v,v_1,v_2}$ (see text).
}
\end{figure}

The {\bf second} mechanism which can induce ripple-rotation is associated with the vectorial terms $\mathbf v_1$ and $\mathbf v_2$.
In Fig.\ref{f.instb} are plotted various solutions of Eq.\ref{Deter} for the branch of $Im(\omega)$ which can admit negative values.
The plot is v.s. $|{\mathbf k}|$ along a given azimuthal direction.
In Fig.\ref{f.instb}a,  we show the weakening and disappearance of the negative minima in a given direction caused by the increase of the component of $\mathbf v_1$ along this direction.
In Fig.\ref{f.instb}b,  we show the re-appearance of the negative minima by inceasing the component of $\mathbf v_2$ along the same direction.
Whereas in Fig.\ref{f.instb}c,  the minima is recovered by decreasing simultaneously the components of both $\mathbf v_1$ and $\mathbf v_2$ along this direction.
In these plots we used: $\gamma = 1$, $\phi = 0.5$, $v = 2$, $\mathcal D = 0.1$, ${\mathcal K}_1 = 0.4$, $s_1 = 7$, ${\mathcal D}_1 = 0.7$ and ${\mathcal D}_2 = 1$. Fig.\ref{f.instb}a varies $v_1$ between 4 and 100 at fixed $v_2=1$.
Fig.\ref{f.instb}b fixes $v_1$ at 100 and increase $v_2$ from 10 to 98.
Fig.\ref{f.instb}c decreases $v_1$ from 100 to 2 and decreases $v_2$ from 1 to 0.02.

An analogous example is given in Fig.\ref{f.2Dinstab} where the position of the minimum rotates by approximately 90 degrees by changing the components $x,y$ of the coefficient $\mathbf{v_1}$.
The sensitivity of the instability to the relative weights of the components of the coefficients $\mathbf{ v_1,v_2}$ along a given direction is rather evident.

The {\bf third} mechanism which influences the orientation of the ripples is associated with the drift velocity $\mathbf v$.
We expect this last mechanism to be more relevant in the case of aeolian dunes and less important in ion-sputtered surfaces.

What we want to stress is that the phase diagram associated with these solutions is very rich and non-trivial.
Instabilities in a given direction can be trigged on and off by changing the relative weights of the quantities $b_x$, $b_y$ or $d_x$, $d_y$ or by varying the intensities or the components of the coefficients $\mathbf{v,v_1,v_2}$.
These changes could be associated with variations of the angular orientation of the ion sputtering.
It is beyond the propose of the present paper to discuss in detail these aspects, however we want to make clear that the present theory can account these phenomena.

%%%%%%%%%%%%%%%%%%%%%%%%%%%%%%%%%%%%%%%%%%%%%%%%%%%%%%%%%%%%%%%%%%%
%%%%%%%%%%%%%%%%%%%%%%%%%%%%%%%%%%%%%%%%%%%%%%%%%%%%%%%%%%%%%%%%%%%
%%%%%%%%%%%%%%%%%%%%%%%%%%%%%%%%%%%%%%%%%%%%%%%%%%%%%%%%%%%%%%%%%%%
%%%%%%%%%%%%%%%%%%%%%%%%%%%%%%%%%%%%%%%%%%%%%%%%%%%%%%%%%%%%%%%%%%%

\section{Saturation/Stabilization} \label{A.stab}

We observed numerically that the behavior of the surface instability in the saturation/stabilization regime described in Section \ref{s.sat} is very sensitive to the presence of a noise term added to Eq.\ref{Gex} (see main text).
In order to investigate this effect we performed several simulations by varying the amplitude ($\sigma$) of an additive Gaussian noise with average zero and variance $\sigma^2$.
Such a parameter $\sigma$ was varied from zero to an upper limit which has been chosen several order of magnitude smaller of the critical roughening $W_c$.
We observe that the stabilization/saturation dynamics is affected by the amplitude of the additive noise.
In particular, in some region of the parameters we observed that the saturation to a constant roughening can be achieved only without any noise contribution, whereas the exponents in the power-law growth appear to be dependent on the amplitude of the noise.
But it must be noted that the observed behavior is extremely complex and shows a strong dependence on the starting configuration, on the history (time steps, random generated noise etc) and on the system parameters. 
After all, this is not surprising in a non-liner system.

Let us here briefly describe the kind of (non-linear) dynamics that we observe numerically in the one-dimensional case.
After the first exponential growth when the Erlich-Schweoebel barrier is active, the system reaches a saturation/stabilization regime where the roughening grows slower than exponential or stays constant.
In the exponential-growth stage we observe that the surface-profile has a typical sinusoidal-like shape (see Fig.\ref{f.saturation}a).
After this stage, at the saturation/stabilization regime the surface tends to assume a triangular like profile flattening in this way its curvature and concentrating it at the vertices (Fig.\ref{f.saturation}b).
In this regime the growth of the surface-ripples might end or -vice versa- it might re-start from small instabilities (infinitesimal noise) generated over the triangular shapes (Fig.\ref{f.saturation}c).
In this second case, the re-nucleated ripples follow a similar history of the ones on which they nucleate: they start with and exponential growth and then reach a saturation until another re-nucleation occurs.
In this phase, interactions among ripples, nucleated in different part of the surface, and interaction between the newly nucleated ripples and the pre-existent surface instabilities plays a major role.
Typically, the overall dynamics appears to be consistent with a power law growth associated with a change in the ripple wavelength, but a finer analysis of this process shows a complex non-uniform growth (see Fig.\ref{f.satNonUnif}).

Figures \ref{f.saturation}a,b were generated by setting: $v = 0.1$, ${\mathcal D} = 0.2$, ${\mathcal K} = 3$, $s=0.6$, $\phi = 10^{-5}$, $\eta = 0.05$, $\gamma = 0.03$, $a =1$, $b=5$, $c=0.1$, $d=0.5$, $\alpha = 10^{5}$, with no additive noise.
Whereas Figs. \ref{f.saturation}c, \ref{f.satNonUnif} have the same set of parameters but with an infinitesimal additive Gaussian noise $\xi$ (Eq.\ref{Gex}) with zero mean and standard deviation equal to $10^{-11}$.

\section{Numerical Solutions} \label{A.sim}

The numerical solutions of Eq.\ref{E1} presented in this paper and in particular the ones shown in Figs.\ref{f.1}, \ref{f.3} and \ref{f.satNonUnif} have been performed as follows.
We considered a one-dimensional flat substrate ($h(x,0)=h0$) of length $L$, with periodic boundary conditions. 
An infinitesimal quantity of mobile atoms were added randomly to the substrate (with $0 < R(x,0) < L/N 10^{-13}$).
We then computed the profile-evolution using Eq.\ref{E1} with the derivative substituted with finite differences. 
To this purpose, the substrate has been divided into $N$ discrete points.
The -adimensional- time indicated in Fig.\ref{f.3} and \ref{f.satNonUnif} is the number of numerical steps.
The height is in unit of $L/N$ and the roughness is defined as $w(t,L)=\left< [h(x',t)- \left< h(x,t) \right>_x]^2 \right>^{1/2}_{x'}$ (see, for instance, \cite{Csa92}).

Several computations with a number of points equal to $N=1000$, 2000 and 3000 (the one presented here have $N=3000$) have been performed to verify the effect of boundary and discretization.
Moreover, simulations with no periodic boundary conditions and with the sputtering term (Eq.\ref{Gex}) applied only to a central mask, have also been performed obtaining very similar results.
The robustness of the present approach has been verified varying the parameters, the time steps, the initial roughness of the substrate, etc.
Comparable results have been always found but, we must stress that, under some conditions, numerical instabilities (in particular surface-deformations with $\lambda \sim L/N$) can be trigged on depending on the protocol utilized.

The simulation result shown in Figs.\ref{f.1}, \ref{f.3} and \ref{f.satNonUnif}  use: $v = 0.1$, ${\mathcal D} = 0.2$, ${\mathcal K} = 3$, $s=0.6$, $\phi = 10^{-5}$, $\eta = 0.05$, $\gamma = 0.03$, $a =1$, $b=5$, $c=0.1$, $d=0.5$, $\alpha = 10^{5}$. 
The simulation time was 30000, steps.

The numerical solutions for the two-dimensional case have been calculated by following the same protocol as in the one-dimensional case.
In two dimensions the substrate is a square $L \times L$ surface with periodic boundary conditions which has been subdivided into a discrete grid of $150 \times 150$ points.
The simulation result shown in Fig.\ref{f.2D} uses: ${\mathbf v} = 0$, ${\mathcal D} = 0$, ${\mathcal K} = 0.5$, $s=0$, $\phi = 0.2$, $\eta = 0.05$, $\gamma = 0.03$, ${\mathbf a} = 0$, $b_x=2$, $b_y=2$ ${\mathbf c}=0$, $d_x=0.3$, $d_y=0.3$, $\alpha = 0$. $b_x=2$, $b_y=5$.
Whereas the simulation in Fig.\ref{f.rotation} uses the same parameters except for $b_x=2$, $b_y=5$ (up) and $b_x=5$, $b_y=2$ (down).
The simulation time was 5000, steps.

%---------------------------------------------
%bibliography
%---------------------------------------------
%
% BibTeX users please use
% \bibliographystyle{}
% \bibliography{}

\begin{thebibliography}{99}
\bibitem{Laplace}
In the case of aeolian-ripples the curvature-dependent terms $b_x$, $b_y$, $d_x$ and $d_y$ are isotropic (equal in the $x$, $y$ directions) and the second derivatives in Eqs.\ref{Gex},\ref{Gad} become proper Laplacians.
\bibitem{time}
The simulation time is in arbitrary units. 
Typically we run between 1000 to 100000 time-steps, depending on the system parameters.
\bibitem{NumIntab}
Numerically we observe the appearance of infinitesimal negative values of $R({\mathbf r},t)$.
If these unphysical negative values are forced to zero, the exponential growth of the roughening interrupts.
\end{thebibliography}
%
% Non-BibTeX users please use

%--------------------------------------------

%\listoffigures

%--------------------------------------------
\end{document}